\documentclass[twocolumn,showpacs,preprintnumbers,amsmath,amssymb]{revtex4}

\usepackage{graphicx}
\usepackage{dcolumn}
\usepackage{bm}

\begin{document}

\preprint{APS/123-QED}

\title{Pressure-Induced Reentrant Oblique Antiferromagnetic Phase in Spin Dimer System TlCuCl$_3$ }

\author{Fumiko Yamada$^1$}
  \email{yamada@lee.phys.titech.ac.jp.}
\author{Yasuyuki Ishii$^2$}
\author{Takao Suzuki$^2$}
\author{Teiichiro Matsuzaki$^2$} 
\author{Hidekazu Tanaka$^1$}
\affiliation{$^1$Department of Physics, Tokyo Institute of Technology, Meguro-ku, Tokyo 152-8551, Japan\\
$^2$Advanced Meson Science Laboratory, Nishina Center, RIKEN, 2-1 Hirosawa,
Wako, Saitama 351-0198, Japan}

\date{\today}

\begin{abstract}
Magnetization measurements under a hydrostatic pressure of $P\,{=}\,1.4$ GPa were performed on the coupled spin dimer system TlCuCl$_3$, which exhibits a pressure-induced quantum phase transition from a gapped singlet state to an antiferromagnetic state at $P_{\rm c}\,{=}\,0.042$ GPa. Antiferromagnetic ordering with ordered moments parallel to the $ac$ plane was observed at $T_{\rm{N}}\,{=}\,16.7$ K at 1.4 GPa. The spin reorientation phase transition was observed at $T_{\rm{R}}\,{\simeq}\,9.2$ K for zero magnetic field, at which the ordered moments start to incline towards the $b$ axis. With increasing external field parallel to the $b$ axis, a second-order phase transition from the oblique antiferromagnetic (OAF) phase to the spin-flop (SF) phase occurs below $T_{\rm{R}}$. We argue that the OAF phase arises from the competition between the anisotropic energies of the conventional second order and the fourth order, that increases with increasing pressure. We discuss the OAF-SF transition within the framework of the mean field approximation.
\end{abstract}

\maketitle


\section{Introduction}
Recently, quantum spin systems composed of antiferromagnetic spin dimers have been attracting considerable attention from the viewpoint of the quantum phase transition (QPT), which is a phase transition between different quantum ground states induced by a continuous change in interaction constants or applied field \cite{Sachdev,Sachdev2}. Such spin dimer systems often have gapped singlet ground states and undergo the QPT to magnetically ordered states upon the application of external magnetic field \cite{Shiramura,O_mag,Nikuni,tanaka_n,Rueegg,O_k,Waki,Jaime,Tsujii,Zapf,Stone,Giamarchi}. This field-induced magnetic ordering can be understood as the Bose Einstein condensation (BEC) of spin triplets called magnons or triplons \cite{Nikuni,Giamarchi2,Wessel,Rice,Matsumoto1,Stone,Kawashima,Misguich,yamada_BEC}.

TlCuCl$_3$ is an $S{=}1/2$ spin dimer system, in which a pair of Cu$^{2+}$ spins in the chemical dimer Cu$_2$Cl$_6$ forms an antiferromagnetic dimer \cite{tanaka_n}. The exchange interactions between neighboring dimers are three-dimensional. The magnitude of the excitation gap is ${\Delta}/k_{\rm B}\,{=}\,7.5$ K \cite{Shiramura,O_mag}. The small gap,  compared with the intradimer interaction $J/k_{\rm B}\,{=}\,65.9$ K, is attributed to large interdimer interactions \cite{Cavadini,Oosawa1}. TlCuCl$_3$ undergoes the magnetic QPT not only in magnetic field but also under hydrostatic pressure \cite{goto_t,Oosawa3,O_n,Rueegg2}. The critical pressures obtained through magnetization measurement and a neutron scattering experiment are $P_{\rm c}\,{=}\,0.042$ GPa \cite{goto_t} and 0.107 GPa \cite{Rueegg2}, respectively. The pressure-induced magnetic QPT is simultaneously caused by the softening and the BEC of the triplet excitations \cite{Matsumoto2,Nohadani2,Rueegg3}. The softening of the triplet excitations results from the decrease in the intradimer exchange interaction and the increase in the interdimer exchange interactions under hydrostatic pressure \cite{goto_m,goto_n}. 
 
\begin{figure}[htbp]
	\begin{center}
		\includegraphics[scale =0.45]{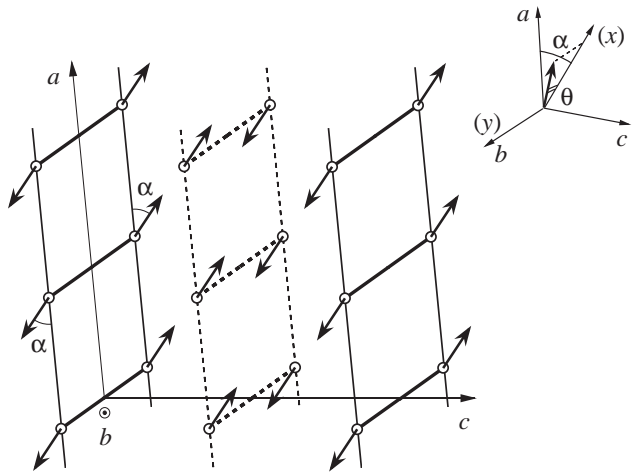}
	\end{center}
	\caption{Spin structure of pressure-induced ordered state in TlCuCl$_3$. The inset shows the definitions of the angles $\alpha$ and ${\theta}$ representing the spin direction. The angle $\theta$ denotes the angle between the ordered spin moment and the $ac$ plane, and the angle $\alpha$ denotes the angle between the $a$ axis and the ordered spin moment projected onto the $ac$ plane.}
	\label{fig:structure}
\end{figure}

When an external magnetic field is applied parallel to the $[2,\,0,\,1]$ direction for $P$\,$>$\,$P_{\rm c}$, the spin-flop transition is observed at $H_{\rm sf}\,{\simeq}\,0.7$ T \cite{goto_t}, which indicates that the spin direction in the pressure-indued ordered state is close to the $[2,\,0,\,1]$ direction.
Neutron elastic scattering experiments performed at $P\,{=}\,1.48$ GPa \cite{Oosawa3,O_n} revealed that the spin direction just below the ordering temperature $T_{\rm N}\,{=}\,16.9$ K is in the $ac$ plane and close to the $[2,\,0,\,1]$ direction, as shown in Fig. \ref{fig:structure}, with ${\theta}\,{=}\,0^{\circ}$ and ${\alpha}\,{\simeq}\,43^{\circ}$. This spin direction is identical to those in the magnetic-field-induced ordered state for $H\,\|\,b$ \cite{tanaka_n} and in the impurity-induced ordered state in TlCu$_{1-x}$Mg$_x$Cl$_3$ \cite{O_i}.

Oosawa {\it et al.} \cite{O_n} performed a polarized neutron elastic scattering experiment at $P\,{=}\,1.48$ GPa and demonstrated that TlCuCl$_3$ undergoes another phase transition at $T_{\rm R}\,{\simeq}\,10$ K, at which the ordered spins lying in the $ac$ plane start to incline toward the $b$ axis. The angle $\theta$ between the ordered spins and the $ac$ plane increases to $40^{\circ}$ with decreasing temperature \cite{O_n}. No such oblique antiferromagnetic (OAF) phase was observed in the previous magnetization measurement for $P$\,$<$\,0.8 GPa \cite{goto_t}. Therefore, the OAF phase is expected to occur for $P$\,$>$\,0.8 GPa. 
To investigate this pressure-induced successive magnetic phase transition and obtain the phase diagram including the OAF phase, we performed the magnetization measurement under  hydrostatic pressure $P\,{=}\,1.4$ GPa. In this paper, we report the results and discuss the mechanism leading to the OAF phase.


\section{Experimental Details}
Single crystals of TlCuCl$_3$ were grown by the vertical Bridgman method. The  details of sample preparation were reported in reference \cite{O_mag}.
The magnetization measurements were performed using SQUID magnetometer (Quantum Design MPMS XL) in the temperature region $2.2\,\mathrm{K}\,{\leq}\,T\,\leq\,50\,\mathrm{K}$ in magnetic fields of up to 1 T. The magnetic fields were applied parallel to the $b$ axis. The hydrostatic pressure $P\,{=}\,1.4$ GPa is applied using a cylindrical high-pressure clamp cell designed for use with the SQUID magnetometer. A sample of size $2.5\,{\times}\,2.5\,{\times}\,4.5$ mm$^3$ was set in the cell with its $b$ axis parallel to the cylindrical axis. As pressure-transmitting fluid, Daphne 7373 oil (Idemitsu Oil \& Gas) was used. The pressure was calibrated with the shift of superconducting transition temperature $T_{\rm c}$ of tin placed in the pressure cell.


\section{Results and Discussions}

\begin{figure}[t]
\includegraphics[scale =0.5]{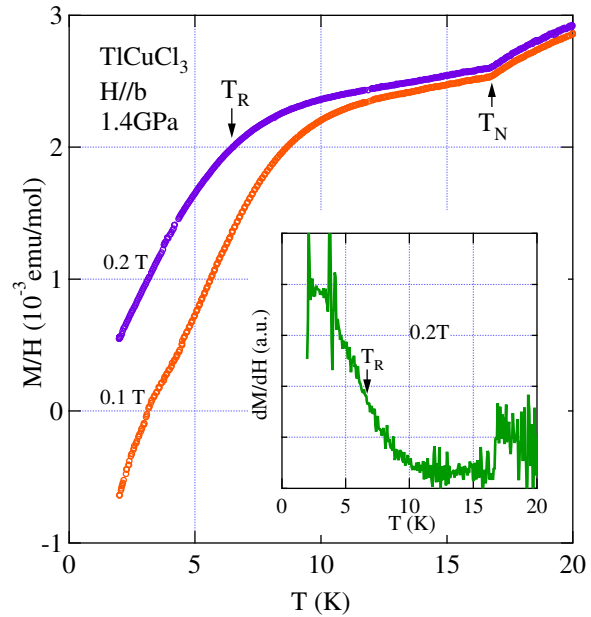}
\caption{Temperature dependence of magnetic susceptibilities in TlCuCl$_3$ measured at $H{=}0.1$ and 0.2 T for $P{=}1.4$ GPa. Inset shows $dM/dT$ vs $T$ measured at $H{=}0.2$ T. 
}
 \label{fig:mt}
 \end{figure}
 
 \begin{figure}[htbp]
	\begin{center}
		\includegraphics[scale =0.60]{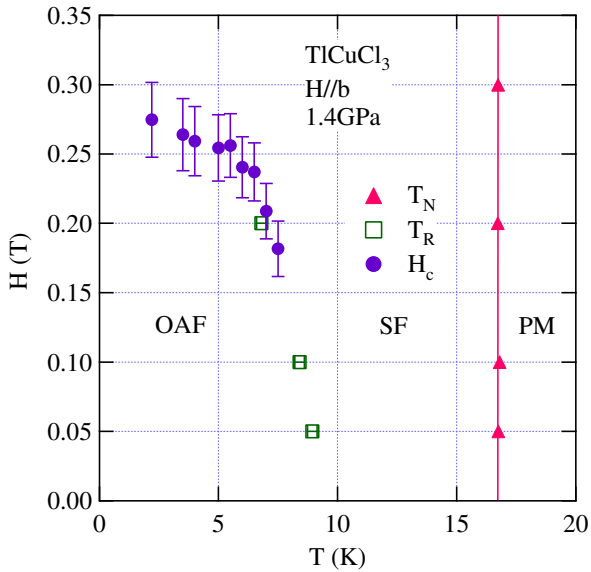}
	\end{center}
	\caption{Phase diagram for magnetic field vs temperature at $P\,{=}\,1.4$ GPa.The magnetic field is applied parallel to the $b$ axis. OAF, SF and PM mean the oblique antiferromagnetic phase, spin-flop phase and  paramagnetic phase, respectively. Triangles and squares denote N\'{e}el temperature $T_{\rm N}$ and spin reorientation transition temperature $T_{\rm R}$, respectively, that are determined from the temperature scan of magnetization. Circles mean the magnetic field $H_{\rm c}$ giving a peak in $dM/dH$ as shown in the inset of Fig. 4.}
	\label{fig:phase}
\end{figure}

\begin{figure}[htbp]
	\begin{center}
		\includegraphics[scale =0.48]{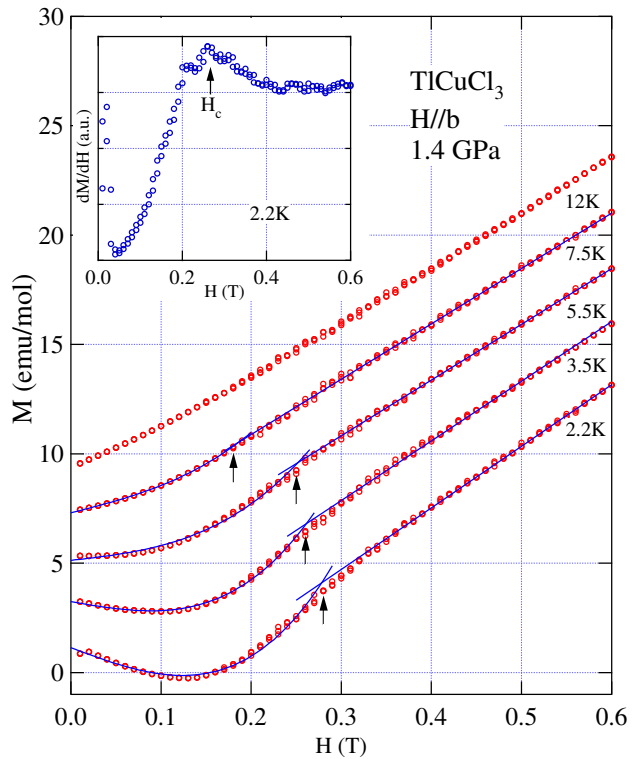}
	\end{center}
	\caption{Magnetization curves of TlCuCl$_3$ for $H\,\|\,b$ measured at various temperatures under hydrostatic pressure $P\,{=}\,1.4$ GPa. Each plot is shifted upward successively by 2 emu/mol for clarity. The solid lines represent magnetization curves  calculated using Eqs.\,(\ref{eq:Mt}) and (\ref{eq:Mv}). Inset shows $dM/dH$ vs $H$ measured at $T{=}2.2$ K.}
	\label{fig:mh}
\end{figure}

As temperature is lowered, the magnetic susceptibility of TlCuCl$_3$ under the hydrostatic pressure of $P\,{=}\,1.4$ GPa exhibits a broad maximum at $T_{\rm max}\,{\simeq}\,35$ K and then decreases, as observed at ambient pressure \cite{O_mag}. Figure\,\ref{fig:mt} shows the low-temperature static susceptibility $\chi\,{=}\,M/H$. The magnetic susceptibility for $H\,{=}\,0.1$ T is negative below 2.5 K. This behavior is unphysical and caused by the diamagnetic background due to the pressure cell. We measured the magnetization of the pressure cell without sample to estimate the diamagnetic background and subtracted its magnetization from the experimental result. However, the correction for the diamagnetic background is insufficient, because the background can not be reproduced completely. 
The magnetic susceptibility displays a bend anomaly at $T_{\rm N}\,{=}\,16.7$ K indicative of antiferromagnetic long range ordering. The N$\acute{\rm e}$el temperature is almost independent of the external magnetic field up to 1 T, and is consistent with $T_{\rm N}\,{=}\,16.9$ K observed in  the polarized neutron elastic scattering experiment under a similar hydrostatic pressure of $P\,{=}\,1.48$ GPa \cite{O_n}. 

The magnetic susceptibility is nearly independent of temperature down to $T_{\rm R}$. This behavior is typical of the perpendicular susceptibility, and thus, the ordered moments lie in the $ac$ plane for $T_{\rm N}$\,$>$\,$T$\,$>$\,$T_{\rm R}$. It is noteworthy that with further decreasing temperature, the magnetic susceptibility decreases rapidly. 
When the ordered moments are inclined by an angle ${\theta}$ from the $ac$ plane (see the inset of Fig.\,1), the magnetic susceptibility of the antiferromagnet composed of two sublattices is given by ${\chi}={\chi}_{\perp}\,{\cos}^2\,{\theta}+{\chi}_{\parallel}\,{\sin}^2\,{\theta}$, where ${\chi}_{\perp}$ is the perpendicular susceptibility that is approximately constant in the ordered phase and ${\chi}_{\parallel}$ is the parallel susceptibility that decreases toward zero for $T\,{\rightarrow}\,0$.
Thus, the rapid decrease in the susceptibility below $T_{\rm R}$ can be attributed to the inclination of the ordered moments toward the external field direction parallel to the $b$ axis, i.e., a spin reorientation transition occurs at $T_{\rm R}$. 

We assign the phase transition temperature $T_{\rm R}$ to the temperature of inflection in the temperature derivative of magnetization $dM/dT$ as shown in the inset of Fig.\,\ref{fig:mt}.
$T_{\rm N}$ and $T_{\rm R}$ obtained from the temperature scan of the magnetic susceptibility are summarized in Fig.\,\ref{fig:phase}. The transition temperature $T_{\rm R}$ decreases as the external magnetic field increases, while the N{\'e}el temperature $T_{\rm N}$ is almost independent of temperature. The transition temperature $T_{\rm R}$ extrapolated to $H\,{=}\,0$ T is $T_{\rm R}\,{\simeq}\,9.2$ K. 
In the previous magnetization measurement for $P$\,$<$\,0.8 GPa \cite{goto_t}, the antiferromagnetic ordering corresponding to $T_{\rm N}$ was observed for $P$\,$>$\,$P_{\rm c}\,{=}\,0.042$ GPa, but no anomaly indicative of the additional transition was observed. Thus, the spin reorientation transition emerges between 0.8 and 1.4 GPa.

Figure\,\ref{fig:mh} shows magnetization curves for $P\,{=}\,1.4$ GPa measured at various temperatures. 
In the data for $T\,{=}\,2.2$ and 3.5 K, there are low-field regions with $dM/dH$$<$ 0. This unphysical behavior is a result of the correction of background due to the pressure cell being  insufficient.
For $T{\leq}7.5$ K, a bend anomaly is observed at the field indicated by arrows in Fig.\,\ref{fig:mh}. We define the transition field $H_{\rm c}$ as the field giving a peak in $dM/dH$, as shown in the inset of Fig.\,\ref{fig:mh}. No hysteresis was observed around $H_{\rm c}$. The transition fields $H_{\rm c}$ obtained at various temperatures are summarized in Fig.\,\ref{fig:phase}. The phase boundaries determined from $H_{\rm c}$ and $T_{\rm R}$ are consistent with each other.
The error bar for $H_{\rm c}$ shown in Fig.\,\ref{fig:phase} corresponds to the width of the peak in $dM/dH$. With increasing temperature, the transition field $H_{\rm c}$ decreases. The anomaly observed at $H\,{=}\,H_{\rm c}$ becomes weak as the temperature increases, and for $T$\,$>$\,7.5 K, the transition field $H_{\rm c}$ cannot be detected. The magnetization at $T\,{=}\,12$ K is almost proportional to external field.  

The ordered moments for $H$\,$>$\,$H_{\rm c}$ lie in the $ac$ plane that is perpendicular to the external field, because the magnetization is proportional to external field, and its slope is almost the same as the magnetic susceptibility for $T_{\rm N}$\,$>$\,$T$\,$>$\,$T_{\rm R}$. The transition at $H_{\rm c}$ is different from the conventional spin-flop transition accompanied with a magnetization jump, because the present transition is continuous. Since there is a certain amount of error in the absolute value of the magnetic susceptibility at low temperatures, the angle between the ordered moments and the $ac$ plane in the ground state cannot be determined from the susceptibility data. However, from the results of neutron elastic scattering experiments\,\cite{Oosawa3,O_n} and the magnetic susceptibility described above, we can deduce that the ordered moments are inclined from the $ac$ plane toward the $b$ axis for $T$\,$<$\,$T_{\rm R}$ and $H$\,$<$\,$H_{\rm c}$, i.e., an oblique antiferromagnetic (OAF) state is realized in the low-temperature and low-field region. As shown below, the continuous phase transition at $H_{\rm c}$ is well described as the transition from the OAF phase to the spin-flop (SF) phase. Here, in the SF phase, the ordered moments are perpendicular to the external magnetic field, i.e., the SF phase is identical to the ordered phase between $T_{\rm N}$ and $T_{\rm R}$. 

Next, we discuss the magnetic-field induced phase transition using the mean-field approximation based on the two-sublattice model. The OAF state cannot be stabilized within the anisotropy of the second order that arises from the dipole-dipole interaction and the anisotropic exchange interaction, because the spin axis is confined to be parallel or perpendicular to the principal axis of the anisotropy. Therefore, we need the anisotropy of  fourth order, as discussed by Igarashi and Nagata\,\cite{Igarashi1,Igarashi2} for the ground state in the mixed antiferromagnet CsMn$_{1-x}$Co$_x$Cl$_3$$\cdot$H$_2$O.

From the the results of magnetization measurements under various pressures\,\cite{goto_t} and ESR measurements at ambient pressure\,\cite{Glazkov,Kolezhuk}, it is expected that the easy axis at low pressure is close to the $[2,\,0,\,1]$ direction (${\alpha}\,{=}\,51^{\circ}$ in Fig.\,1) and the $b$ axis is the second easy axis. We define the $x$ and $y$ axes to be parallel to the easy axis in the $ac$ plane and the $b$ axis, respectively, as shown in the inset of Fig.\,1, and assume that the ordered moments at $P\,{=}\,1.4$ GPa are restricted to the $xy$ plane. We introduce the phenomenological anisotropy energy of fourth order to stabilize the OAF state and then write the energy of the system as
\begin{eqnarray}
E&=&-\left({\bm M}_1+{\bm M}_2\right){\cdot}{\bm H}+A{\bm M}_1{\cdot}{\bm M}_2 \nonumber \\ 
&&-\frac{K}{2}M_0^2\left({\beta}_1^2+{\beta}_2^2\right)+\frac{L}{4}M_0^4\left({\beta}_1^4+{\beta}_2^4\right)\,,
\label{eq:E1}
\end{eqnarray} 
where ${\bm M}_1$ and ${\bm M}_2$ are sublattice magnetizations and $M_0$ denotes their magnitude. The first and second terms of eq.\,(\ref{eq:E1}) are the Zeeman term and the exchange energy, respectively. The third and last terms are anisotropic energies of the second order and the fourth order, respectively. ${\beta}_1$ and ${\beta}_2$ are direction cosines of ${\bm M}_1$ and ${\bm M}_2$ to the $x$ axis, respectively, and coefficients $K$ and $L$ are both positive. The second-order anisotropy prefers ordered moments pointing in the $x$ direction, while the forth order anisotropy prevents such moments. The condition $AM_0\,{\gg}\,H, KM_0, LM_0^3$ is satisfied in the present case.

 \begin{figure}[htbp]
	\begin{center}
		\includegraphics[scale =0.90]{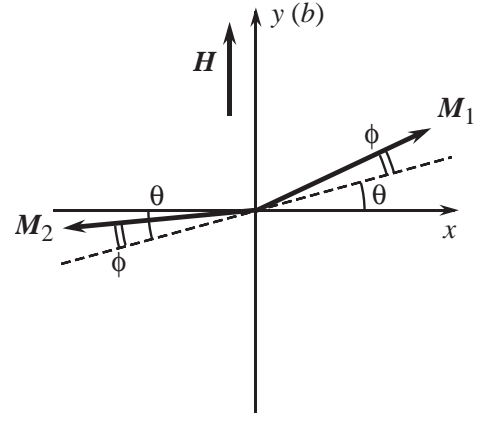}
	\end{center}
	\caption{Configuration of sublattice magnetizations ${\bm M}_1$ and ${\bm M}_2$. Dashed line denotes the spin axis that is canted from the $x$ axis by angle $\theta$.
	}
	\label{fig:SubMag}
\end{figure}

The sublattice magnetizations in the ground state are expressed as
\begin{eqnarray}
{\bm M}_1&=&M_0\left({\cos}\,({\theta}+{\phi}), {\sin}\,({\theta}+{\phi}), 0\right)\,,\nonumber \\
{\bm M}_2&=&M_0\left(-{\cos}\,({\theta}-{\phi}), -{\sin}\,({\theta}-{\phi}), 0\right)\,,
\label{eq:SubMag}
\end{eqnarray} 
where angles of $\theta$ and $\phi$ are defined in Fig.\,\ref{fig:SubMag}. ${\beta}_1$ and ${\beta}_2$ in eq.\,(\ref{eq:E1}) are given by ${\beta}_1={\cos}\,({\theta}+{\phi})$ and ${\beta}_2={\cos}\,({\theta}-{\phi})$. Substituting eq.\,(\ref{eq:SubMag}) into eq.\,(\ref{eq:E1}), the equilibrium conditions are given as
\begin{eqnarray}
{\sin}\,{\phi}=\frac{H}{2AM_0}{\cos}\,{\theta}\,,
\label{eq:phi}
\end{eqnarray}
and
\begin{eqnarray}
\cos^2\theta &=& \frac{H^2 + 2AKM_0^2}{2ALM_0^4}\,,
\label{eq:theta}
\end{eqnarray} 
for $0\,{\leq}\,H\,{\leq}\,H_{\rm c}$ and 
$\theta = 0^\circ$
for $H$\,$>$\,$H_{\rm c}$, where $H_{\rm c}$ is the transition field given by
\begin{eqnarray}
H_{\rm c} = \sqrt{2A(LM_0^2 - K)}M_0\,,
 \label{eq:Hc}
\end{eqnarray} 
which is obtained by setting ${\theta}\,{=}\,0^{\circ}$ in eq.\,(\ref{eq:theta})\,.

\begin{figure}[htbp]
	\begin{center}
		\includegraphics[scale =0.52]{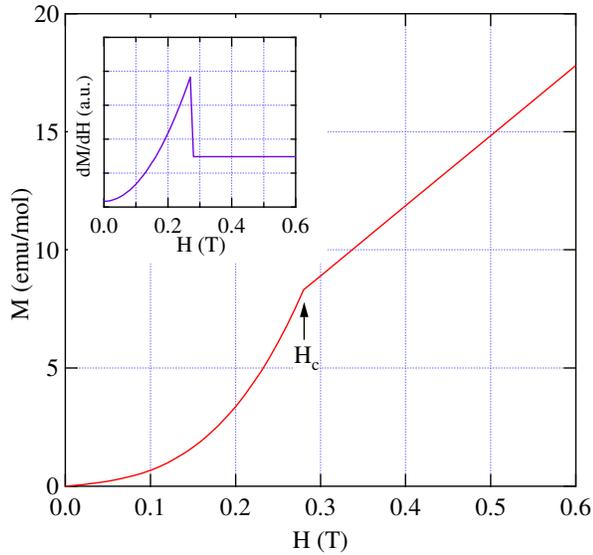}
	\end{center}
	\caption{Calculated magnetization curves and its field derivative $dM/dH$ with  $H_{\rm c}\,{=}\,0.28$ T, ${\theta}_0\,{=}70^{\circ}$ and ${\chi}_{\bot}\,{=}\,3.0\,{\times}\,10^{-3}$ emu/mol.
	}
	\label{fig:m_cal}
\end{figure}

When the condition $LM_0^2$\,$>$\,$K$ is satisfied, the OAF state is stabilized as the ground state. The oblique angle at zero magnetic field is given by ${\cos}^2{\theta}_0\,{=}\,K/(LM_0^2)$.
In this case, the magnetization $M$ is expressed as
 \begin{eqnarray}
 M = \frac{H}{A}\cos^2{\theta}\,.
 \label{eq:M}
\end{eqnarray} 
From eqs.\,(\ref{eq:phi}), (\ref{eq:theta}) and (\ref{eq:M}), we obtain
 \begin{eqnarray}
M = \frac{{\sin}^2{\theta}_0}{AH_{\rm c}^2}H^3 + \frac{{\cos}^2{\theta}_0}{A}H\, 
\label{eq:Mt}
\end{eqnarray} 
for $0\,{\leq}\,H\,{\leq}\,H_{\rm c}$ and 
\begin{eqnarray}
M = \frac{H}{A}={\chi}_{\bot}H\,        
\label{eq:Mv}
\end{eqnarray} 
for $H$\,$>$\,$H_{\rm c}$. A continuous phase transition without any magnetization jump occurs at $H_{\rm c}$. Figure\,\ref{fig:m_cal} shows examples of the magnetization curve and its field derivative $dM/dH$ calculated with $H_{\rm c}\,{=}\,0.28$ T, ${\theta}_0\,{=}70^{\circ}$ and ${\chi}_{\bot}\,{=}\,3.0\,{\times}\,10^{-3}$ emu/mol. The calculated magnetization and $dM/dH$ curves reproduce the features of those plotted in Fig.\,\ref{fig:mh}.

 \begin{figure}[htbp]
	\begin{center}
		\includegraphics[scale =0.65]{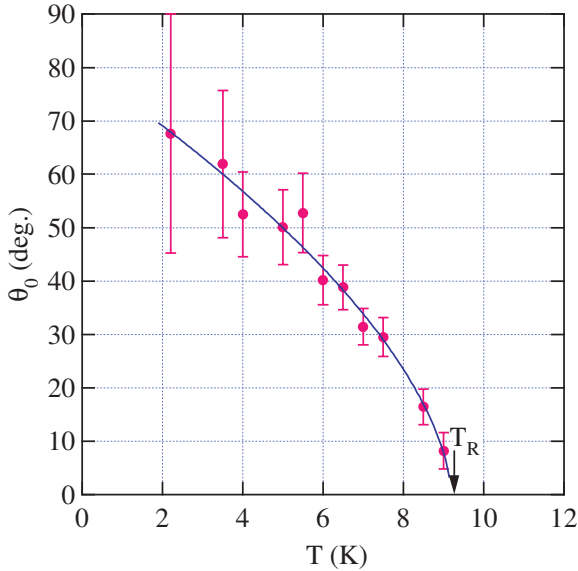}
	\end{center}
	\caption{Temperature dependence of the oblique angle at zero magnetic field $\theta_0$ between the ordered spin moments and the $b$ axis obtained by the present analysis. The solid line is a guide for the eyes.
	}
	\label{fig:theta}
\end{figure}

We analyze the magnetization, using  Eqs.\,(\ref{eq:Hc}), (\ref{eq:Mt}) and (\ref{eq:Mv}), with the remaining background term due to the pressure cell being linear in $H$. The value of $A\,{=}\,1/{\chi}_{\bot}$ is obtained from the slope of the magnetization curve for $H$\,$>$\,$H_{\rm c}$. Solid lines in Fig.\,\ref{fig:mh} are calculated magnetization curves (for 2.2 K, $H_{\rm c}\,{=}\,0.27$ T, ${\theta}_0\,{=}67.7^{\circ}$ and ${\chi}_{\bot}\,{=}\,3.0\,{\times}\,10^{-3}$ emu/mol). The calculated magnetization curves are in good agreement with the experimental result. Although the present analysis is based on the spin configuration of the ground state, i.e., $|{\bm M}_1|\,{=}\,|{\bm M}_2|$ is assumed, the results of the present analysis are applicable at finite temperatures as long as ${\chi}_{\perp}\,{\gg}\,{\chi}_{\parallel}$ is satisfied. This condition should be satisfied for $T$\,$<$\,$T_{\rm N}/3$. Figure\,\ref{fig:theta} shows the temperature dependence of the oblique angle at zero magnetic field ${\theta}_0$. With increasing temperature, the oblique angle ${\theta}_0$ decreases and becomes zero at $T_{\rm R}\,{\simeq}\,9.2$ K. This is because the magnitude of the sublattice magnetization $M_0$ decreases with increasing temperature, and $LM_0^2\,{=}\,K$ at $T_{\rm R}$. At $T\,{=}\,4.0$ K, we obtain ${\theta}_0\,{=}52^{\circ}$, which is somewhat larger than ${\theta}_0\,{=}40^{\circ}$ obtained in the polarized neutron elastic scattering experiment \cite{O_n}. 

In the coupled spin dimer system TlCuCl$_3$, the magnitude of the sublattice magnetization $M_0$ depends on the applied pressure. At $T\,{=}\,0$, $M_0$ is approximately proportional to $(P-P_{\rm c})^{1/2}$ \cite{Matsumoto2}. In the present pressure range of 1 GPa, which is much larger than the critical pressure $P_{\rm c}\,{=}\,0.042$ GPa, the pressure dependence of $M_0$ is given by $M_0^2\,{\simeq}\,{\xi}P$ with coefficient ${\xi}$. Therefore, the pressure dependences of the transition field and oblique angle respectively are 
\begin{eqnarray}
H_{\rm c} = \sqrt{2A{\xi}P(L{\xi}P - K)}\,, 
\label{eq:HcP}
\end{eqnarray}
and
\begin{eqnarray}
{\cos}^2{\theta}_0=K/(L{\xi}P)\,.
\label{eq:theta0P}
\end{eqnarray}
When $P\,{=}\,K/(L{\xi})$ is satisfied with increasing pressure, the OAF phase emerges as the ground state. If the coefficients $K$ and $L$ are independent of temperature, the threshold pressure $P_{\rm OAF}$ of the OAF phase is given by
\begin{eqnarray}
P_{\rm OAF}=P_0\,{\cos}^2{\Theta}_0,
\label{eq:POAF}
\end{eqnarray}
where ${\Theta}_0$ is the oblique angle at $T{=}0$ for the present pressure $P_0{=}1.4$ GPa. Substituting ${\Theta}_0{\simeq}75^{\circ}$ for $T{\rightarrow}0$, we obtain $P_{\rm OAF}{\simeq}0.1$ GPa. This $P_{\rm OAF}$ value is too small, because the threshold of emergence of the OAF phase in TlCuCl$_3$ is between 0.8 and 1.4 GPa. Therefore, we infer that similarly to the dominant intradimer exchange interaction\,\cite{goto_m,goto_n}, the coefficient $K$ for the second-order anisotropy decreases with increasing pressure. To obtain the pressure dependences of the transition field and oblique angle, and also the pressure-temperature phase diagram, detailed magnetization measurements under various hydrostatic pressures are now in progress.


\section{Conclusions}
We have presented the results of magnetization measurements on the gapped spin system TlCuCl$_3$ under a hydrostatic pressure of 1.4 GPa in magnetic fields parallel to the $b$ axis. The pressure-induced antiferromagnetic ordering occurs at $T_{\rm N}\,{=}\,16.7$ K. 
An additional phase transition to the oblique antiferromagnetic (OAF) phase was observed at $T_{\rm R}\,{\simeq}\,9.2$ K and zero magnetic field. This result verifies the observation of polarized neutron scattering performed at 1.48 GPa by Oosawa {\it et al.} \cite{O_n}. With increasing external field, the second order phase transition from the OAF phase to the spin-flop (SF) phase, which is identical to the ordered phase between $T_{\rm N}$ and $T_{\rm R}$,  occurs for $T$\,$<$\,$T_{\rm R}$. We introduced the phenomenological anisotropic energy of  fourth order, that prevents the ordered moments from lying in the $ac$ plane and discussed the ground state using the mean-field approximation based on the two-sublattice model. Our model describes (i) the emergence of the OAF state with increasing pressure, (ii) the decrease of the oblique angle toward zero at $T_{\rm R}$ with increasing temperature, and (iii) magnetization curve associated with the continuous OAF$-$SF phase transition. The origin of the fourth-order anisotropy may be the magnetoelastic coupling and the coupling between applied pressure and linear strain. However, definite mechanism leading to the forth-order anisotropy is an open question.


\begin{acknowledgments}
We express our sincere thanks to A. Oosawa for fruitful discussion. This work was supported by a Grant-in-Aid for Scientific Research (A) from the Japan Society for the Promotion of Science (JSPS) and the Global COE Program ``Nanoscience and Quantum Physics''  at Tokyo Tech funded by the Japanese Ministry of Education, Culture, Sports, Science and Technology. T. S. is supported by a Grant-in-Aid for Young Scientists (B) 19740187 from JSPS. 
\end{acknowledgments}


\end{document}